# Superconductivity in a layered cobalt oxychalcogenide $Na_2CoSe_2O$ with a triangular lattice

Jingwen Cheng[1,2,§], Jianli Bai[1,2,§], Binbin Ruan[1], Pinyu Liu[1,2], Yu Huang[1,2], Qingxin Dong[1], Yifei Huang[1], Yingrui Sun[1,2], Cundong Li[1,2], Libo Zhang[1,2], Qiaoyu Liu[1,2], Wenliang Zhu[1], Zhian Ren[1,2], and Genfu Chen[1,2,3,*]

[1]Institute of Physics and Beijing National Laboratory for Condensed Matter Physics, Chinese Academy of Sciences, Beijing, 100190, China

[2]School of Physical Sciences, University of Chinese Academy of Sciences, Beijing, 100049, China

[3]Songshan Lake Materials Laboratory, Dongguan, Guangdong 523808, China

**ABSTRACT**

Unconventional superconductivity in bulk materials under ambient pressure is extremely rare among the 3$d$ transition-metal compounds outside the layered cuprates and iron-based family [1-2]. It is predominantly linked to highly anisotropic electronic properties and quasi-two-dimensional (2D) Fermi surfaces. To date, the only known example of the Co-based exotic superconductor was the hydrated layered cobaltate, $Na_xCoO_2 \cdot yH_2O$, and its superconductivity is realized in the vicinity of a spin-1/2 Mott state [3-5]. However, the nature of the superconductivity in these materials is still a subject of intense debate, and therefore, finding new class of superconductors will help unravel the mysteries of their unconventional superconductivity. Here we report the discovery of superconductivity at ~ 6.3 K in our newly synthesized layered compound $Na_2CoSe_2O$, in which the edge-shared $CoSe_6$ octahedra form [$CoSe_2$] layers with a perfect triangular lattice of Co ions. It is the first 3$d$ transition-metal oxychalcogenide superconductor with distinct structural and chemical characteristics. Despite its relatively low $T_C$, this material exhibits very high superconducting upper critical fields, $\mu_0H_{C2}(0)$, which far exceeds the Pauli paramagnetic limit by a factor of 3 - 4. First-principles calculations show that $Na_2CoSe_2O$ is a rare example of negative charge transfer superconductor. This cobalt oxychalcogenide with a geometrical frustration among Co spins, shows great potential as a highly appealing candidate for the realization of unconventional and/or high-$T_C$ superconductivity beyond the well-established Cu- and Fe-based superconductor families, and opened a new field in physics and chemistry of low-dimensional superconductors.

**INTRODUCTION**

After the discovery of superconductivity with high-$T_C$ in iron-based family [2], extensive work has been conducted on cobalt analogs, $LnCoPnO$ ($Ln$ = lanthanoids; $Pn$ = P or As), $AeCo_2Pn_2$ ($Ae$ = alkali earths), and $ACo_2Ch_2$ ($A$ = alkali metal, $Ch$ = S, Se, Te) to understand their physical properties and the origin of superconductivity in iron-based superconductors, since the cobalt is next to iron and has just one more electron than iron [6-8]. Although these compounds have crystal structures and Fermi surfaces similar to their iron-based counterparts, and their magnetism was largely tuned by size and electronic effects from changing the interlayer spacing of [$CoPn$] or [$CoCh$]. Unfortunately, no superconductivity has been reported so far in such tetragonal Co-based materials. On the other hand, there is a hydrated layered cobaltate, $Na_xCoO_2·yH_2O$, where the cobalt ions ($S$ = 1/2) lie on a triangular lattice, which was discovered to exhibit superconductivity at around 5 K two decades ago [3]. Superconductivity on such 2D triangular lattice [$CoO_2$] layer has attracted much attention, because it may stabilize novel resonating valence bond (RVB) states [4], which was not realized in copper oxides with the 2D square lattice. Remarkably, a chiral $d + id$ -wave topological superconducting state was theoretically proposed for this material [9]. The vicinity of strongly correlated properties and frustrated magnetism, would make this material the most exotic superconductor discovered so far.

However, the realization of high-quality samples of $Na_xCoO_2·yH_2O$ is still a big challenge to date, in which superconductivity occurs only in the low Na content region ($x = 0.22 \sim 0.47$), and when water is incorporated into its structure [10-12]. The control of the Na composition and the amount of water are the two key factors limiting the obtaining of a single phase of $Na_xCoO_2 \cdot yH_2O$ with good superconducting properties. Furthermore, this compound is chemically unstable at ambient temperature and humidity, which makes its handling and characterization problematic [12]. Accordingly, the nature of the superconductivity in this material is still under debate due to the lack of well-reproducible and reliable experimental results. Hence, exploring its possible analogs will deepen our understanding of the hydrated sodium cobaltate superconductor physics.

In this work, we have successfully synthesized a new cobalt oxychalcogenide, i.e., $Na_2CoSe_2O$, in the Na-Co-Se-O system. This compound has a layered structure composed of alternate stacking of edge-sharing $Na_6O$ octahedra and $CoSe_6$ octahedra, in which Co atoms form a 2D triangular lattice, analogous to the hydrated sodium cobaltate, $Na_xCoO_2 \cdot yH_2O$, while the $CoSe_2$ layer acts as the conducting layer instead of $[CoO_2]$ layer, and $[Na_2O]$ layer instead of insulating layer of $[Na_x(H_2O)_y]$. Superconductivity was observed with $T_C$ up to 6.3 K in the compound. Like the $[CuO_2]$ plane in the cuprates and the $[Fe_2Pc_2]$ (*Pc*: anions of P, As, S, Se or Te) layer in the iron-based family, the geometric frustrated cobalt dichalcogenide layer, $[CoSe_2]$, is believed crucial to support superconductivity in such Co-based oxychalcogenide superconductor. First-principles calculations show that states on Fermi level ($E_F$) are mainly contributed by the O-2$p$ orbitals, while the Co-3$d$ states dominate below –1 eV. $Na_2CoSe_2O$ serves as a rare example of negative charge transfer superconductor. Based on the flexibility of the structure, new Co-based superconductors with various stacking structures can be designed by changing the blocking layers. Our results demonstrate a new novel family of superconductors with unique structural and chemical characteristics, and offer new perspectives for the possible further raising of the superconducting transition temperature.

**EXPERIMENTAL SECTION**

**Crystal Growth.** Single crystals of $Na_2CoSe_2O$ were produced by a solid-state reaction of $Na_2O_2$ (95%, shot), NaSe, and CoSe. NaSe was synthesized in liquid ammonia from Na (2N5, lump) and Se (5N, shot) in stoichiometric quantities. CoSe were prepared from elemental Co (2N8, powder) and Se at 600 °C for 20 h. These reactants (in total 3g) were then mixed in stoichiometric ratios, pressed, and put into an alumina crucible, and finally sealed in a quartz tube filled with Ar under a pressure of 0.4 atmosphere. The quartz tube was heated to 650 °C within 20 h, held there for 20 h, and then cooled to 450 °C over 100 h. The as grown flaky single crystals are uniformly distributed throughout the reaction products with dimensions 0.3×0.5×0.02 $mm^3$ and can be separated mechanically. The crystals are unstable in moist air. Except for the heat treatment, all the material preparation procedures were carried out in an Ar-filled glove box ($O_2$, $H_2O$ < 1 ppm).

**Structure and Composition Determination.** Structure analyses were done on a Single Crystal Diffractometer of BRUKER D8 VENTURE. A specimen of $Na_2CoSe_2O$, approximate dimensions 0.050 mm × 0.270 mm × 0.340 mm, was used for the X-ray crystallographic analysis and protected under Ar atmosphere during data collection. A total of 1094 frames were collected. The frames were integrated with the Bruker SAINT software package using a narrow-frame algorithm. The integration of the data using a trigonal unit cell yielded a total of 869 reflections to a maximum θ angle of 26.77° (0.79 Å resolution), of which 112 were independent (average redundancy 7.759, completeness = 96.6%, $R_{int}$ = 5.53%, $R_{sig}$ = 3.17%) and 110 (98.21%) were greater than $2\sigma(F^2)$. The final cell constants are based upon the refinement of the XYZ-centroids of 1108 reflections above 20 σ(I) with 8.507° < 2θ < 56.33°. The ratio of minimum to maximum apparent transmission was 0.456. The structure was solved and refined using the Bruker SHELXTL Software Package [13] (more details in Supporting Information).

The chemical compositions were checked with an energy dispersive X-ray (EDX) spectrometer equipped on a Phenom scanning electron microscope (SEM) operated at

15 kV. The EDX analysis was performed on the fresh surface cleaved from as grown crystals. EDX analysis confirmed the existence of four elements Na, Co, Se and O, with Na : Co : Se close to stoichiometric proportions (more details in Supporting Information).

**Physical Property Measurements.** Resistivity and Hall coefficient measurements were performed in a Quantum Design physical property measurement system (PPMS). Standard four-probe and Hall-bar method were adopted to obtain the resistivity and Hall coefficient, respectively. The magnetization was measured by a Quantum Design magnetic property measurement system (MPMS). All samples used for measurements were protected by N-Grease to avoid deteriorating in the air.

**Theoretical Methods.** First-principles calculations were performed with the density functional theory (DFT), as implemented in the QUANTUM ESPRESSO package [14], using experimental crystallographic data. To address the Co-3$d$ electrons, the DFT + $U$ method was applied with the Hubbard $U$ parameter set to 4 eV for Co. The projector augmented wave pseudopotentials of PBEsol were applied [15-16]. Monkhorst–Pack grids of $17^3$ and $25^3$ were used in the calculations of charge density and density of states (DOS), respectively. The CRITIC2 package was used to perform the Bader analysis [17]. No magnetic order or spin-orbit coupling effects was considered in the calculations.

## RESULTS AND DISCUSSION

$Na_2CoSe_2O$ crystallizes in trigonal space group $R$-$3m$ (No. 166), as shown in Fig.1(a). The lattice parameters are determined to be $a$ = 3.5161(9) Å, $c$ = 28.745(11) Å, and V = 307.8(20) Å$^3$, respectively The sufficiently large ratio, $c/a \approx 8.2$，combining the significant spatial separation between [$CoSe_2$] layers, 9.581(67) Å, are likely to produce 2D effects in its physical properties. In fact, the weak interlayer interaction allows it to form quasi-2D flaky crystals which are easily cleaved. To the best of our knowledge, this new crystal structure is different from the existing reported ones. In general, most of transition metal layered mixed-anion compounds crystallize in tetragonal space groups rather than hexagonal space group. Although the main features of $Na_2CoSe_2O$, namely [$CoSe_2$] layers, have been described in a series of transition-metal dichalcogenides (TMDs) [18], such as $VSe_2$, $NbSe_2$, etc., the insulating [$Na_2O$] layers with triangular lattice between adjacent [$CoSe_2$] layers have never been found in any other compounds. Specifically, it is staggered by three [$Na_2O$] layers and three [$CoSe_2$] layers with ABC-stacking order in one unit cell as shown in Fig. 1(a). [$Na_2O$] layers stack along (1/3, 2/3, -1/3), and [$CoSe_2$] layers stack along (2/3, 1/3, 1/3), respectively. The distance between adjacent [$Na_2O$] and [$CoSe_2$] layers is 4.79083 Å. Six Se (Na) atoms surround a Co (O) atom to form a $CoSe_6$ ($ONa_6$) octahedra. For example, in [$CoSe_2$] layer, Co atoms locate in the same plane, arranged in a triangular lattice and Se atoms alternate locate above and below the center of the triangular lattice with the shortest Co-Co distance of 3.5161(9) Å and Co-Se bond length of 2.3994(12) Å. The nearest distance between Se atoms is also 3.5161(9) Å and the Se-Co-Se angle is 94.23(6)° in the same Se-plane, while 3.266(4) Å and 85.77(6)° in different planes, forming an $CoSe_6$ octahedra compressed along the c-axis. These edge-shared $CoSe_6$ octahedra form the [$CoSe_2$] layers, as shown in Fig. 1(b), (c). Thus, the effect of geometric frustration caused by the triangular arrangement of magnetic atoms can be expected. In the past few decades, geometric frustration has been the subject of a number of studies because it may lead to various exotic ground states, including spin liquid and nematic phases, especially the likely physical realization of a long-sought RVB state in triangular magnetic lattices with antiferromagnetic nearest-neighbor interactions [19-21]. In general, both the [$Na_2O$] and [$CoSe_2$] layers are expected to be electrically neutral, and the charge-neutral nature of the [$Na_2O$] and [$CoSe_2$] layers, in principle, prevents charge transfer between them, thus the ionic state of cobalt in $Na_2CoSe_2O$ might be $Co^{4+}$ and its outer-shell electron is $3d^5$. And so, interesting phenomena related to the geometric frustration, if the nearest-neighbor exchange interactions are antiferromagnetic, may also emerge. Indeed, as shown latter, magnetic susceptibility measurement does show no long range magnetic ordering down to 6 K, at which the superconductivity sets in. $Na_2CoSe_2O$ could thus represent a model system for studying the superconducting state in a geometrically frustrated system.

It is well known that, the hydrated sodium cobaltate, $Na_xCoO_2·yH_2O$, is also a layered crystal structure, possessing [$CoO_2$] layers consisting of $CoO_6$ octahedra, where the [$CoO_2$] layers are separated by thick insulating layers of $Na^+$ ions and $H_2O$ molecules. Remarkably, the atomic arrangement in essential layers, namely, the triangular lattice of Co atoms is the same in [$CoSe_2$] layers and [$CoO_2$] layers, which is the most interesting and attractive structural feature that may lead to the presence of geometric magnetic frustration. As the key structure unit for superconductivity, the geometric frustrated [$CoSe_2$] layers can be simply constructed by substituting Se atoms for O atoms in [$CoO_2$] layers. Besides, [$Na_2O$] layers replacing the insulating layers of [$Na_x(H_2O)_y$] makes the layer spacing between [$CoSe_2$] layers, 9.581(67) Å, in $Na_2CoSe_2O$ comparable to the distance of 9.8103(57) Å between [$CoO_2$] layers in $Na_xCoO_2·yH_2O$. However, compared to $Na_xCoO_2·yH_2O$, the crystals of which must be obtained by extra chemical oxidation process from a parent compound and the Na content of which is difficult to control and identify, it is relatively easy to prepare single crystals of $Na_2CoSe_2O$ with well-defined structure and stoichiometry by solid-state method only, which will be conducive to our understanding of superconducting mechanisms of the geometric frustrated systems.

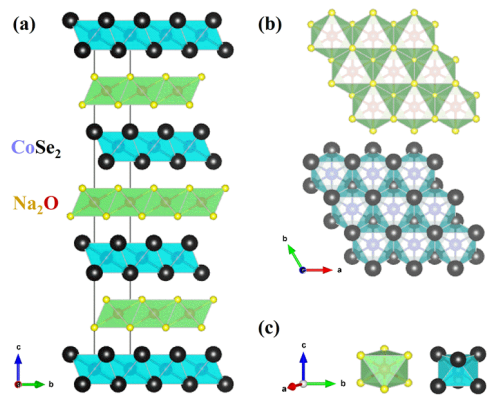

**Fig. 1. Crystal structure of $Na_2CoSe_2O$.** (a) The structure of $Na_2CoSe_2O$, viewing from the direction of *a*-axis. Yellow, red, blue and black spheres represent Na, O, Co and Se atoms,

respectively. (b) The structures of [Na₂O] layer and [CoSe₂] layer. (c) The structures of single Na₆O octahedron and CoSe₆ octahedron.

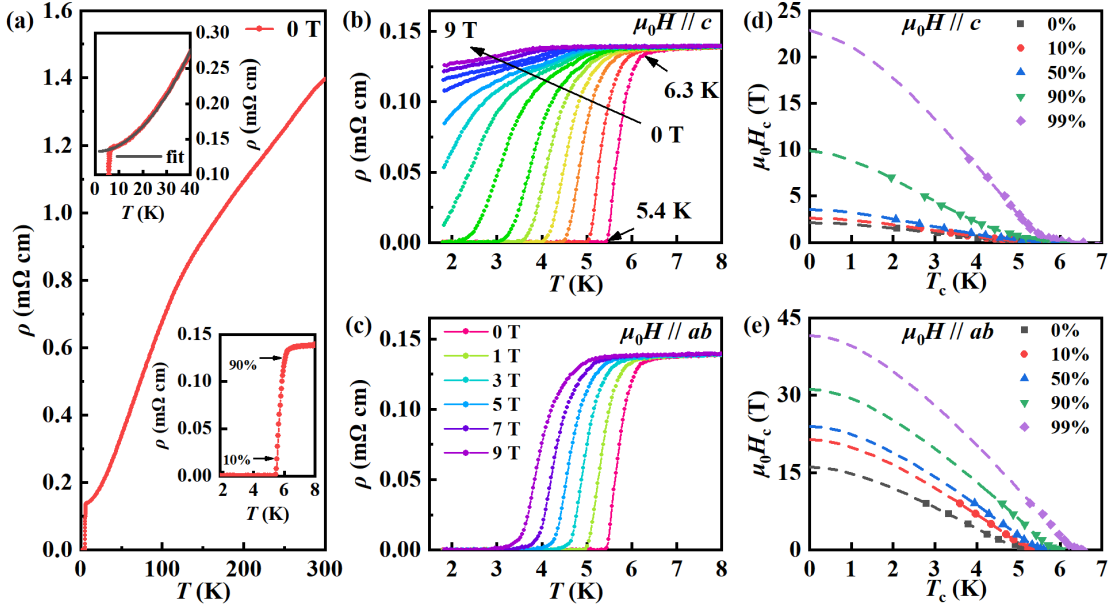

**Fig. 2. Superconductivity and anisotropic upper critical fields of single crystal Na$_2$CoSe$_2$O.** (a) Temperature dependence of the in-plane resistivity of Na$_2$CoSe$_2$O at zero magnetic field. Upper inset: Solid line is the fitting $\rho(T) = \rho_0 + AT^2$ with $\rho_0$ = 0.132 mΩ cm and $A = 8.8 \times 10^{-5}$ mΩ cm/K² for $T_C < T < 40$ K. Lower inset: The enlarged low-temperature data, the arrows mark the superconducting transition temperature with the 90% and 10% criterions. (b) Low-temperature dependence of the resistivity across the superconducting transition with varying magnetic fields from 0 to 9 T for $\mu_0 H \parallel c$, ($\mu_0 H$ = 0, 0.1, 0.3, 0.5, 0.75, 1, 1.5, 2, 2.5, 3, 4, 5, 7 and 9 T). (c) Low-temperature dependence of the resistivity across the superconducting transition with varying magnetic fields from 0 to 9 T for $\mu_0 H \parallel ab$. (d) and (e) Temperature dependence of the upper critical fields $\mu_0 H(T)$ for both directions. The dashed lines are the two-band fit (see the Supporting Information).

Figure 2(a) shows the temperature dependence of in-plane resistivity. The normal-state resistivity is of the order of 1.4 mΩ·cm at room temperature, and shows an anomalous S-shaped temperature dependence, a character of a highly correlated electron system. As a hallmark of the Fermi liquid state formed by strong electron correlations, $\rho(T)$ exhibits also a quadratic temperature dependence over a wide low temperature range ($T_C < T < 40$ K), as shown in the upper inset of Fig. 2(a). The lower inset shows the resistive superconducting transition in more detail. The onset temperature is 6.3 K, with the 10 – 90 % transition width ~ 0.5 K, indicating the high quality of the sample. The upper critical field $H_{C2}$ is one of the important parameters to characterize superconductivity. To get information about $H_{C2}$ of Na$_2$CoSe$_2$O sample, we measured the electrical resistivity under several selected magnetic fields up to 9 T, as shown in Fig. 2(b,c) with the field applied parallel to the *c*-axis ($\mu_0H // c$) and *ab*-plane ($\mu_0H // ab$), respectively. For $\mu_0H // c$, with increasing the field, the transition temperature $T_C$ shifts slowly to lower temperature and the transition width gradually becomes broader. One can see that the onset $T_C$ moves very little (~ 2.5 K) at the field as high as 9 T, while the foot of the transition is shifted to substantially lower temperature. Surprisingly, in contrast to the strong broaden effect observed for $\mu_0H // c$, one can find the superconducting transition is broadened very slightly in magnetic fields up to 9 T for $\mu_0H // ab$. These features are very similar to the high-$T_C$ copper-oxide superconductors, in which the anisotropy of broadening phenomena is related to the orientation of the field with respect to the Cu-O planes because of the strongly anisotropic critical fields and anisotropic pinning forces [22-23]. In our case, this is also actually understandable because the spacing distance between [CoSe$_2$] planes is very large, making the system closer to 2D as well, as expected from its quasi-2D electronic structure (see the inset of Fig. 4).

Figure 2(d,e) shows $H_{C2}$ - $T_C$ curves for both $\mu_0H // c$ and $\mu_0H // ab$, respectively, in which we took five values of transition temperature corresponding to the resistivity drops to 99%, 90%, 50%, 10% of the normal state value and zero resistivity. For each criterion of superconducting transition, $H_{C2}(T)$ lines actually display a very strong upward curvature close to $T_C$, which deviate from the conventional one-band Werthamer-Helfand-Hohenberg (WHH) model [24], and cannot be explained by the Ginzburg-Landau (GL) theory, too [25]. As for the upward curvature of $H_{C2}(T)$, several theoretical approaches have been proposed, like the mesoscopic fluctuation in disordered superconductors [26], the Josephson tunneling between superconducting clusters [27], the reduction of the diamagnetic pair-breaking in the stripe phase [28], the strong spin-flip scattering [29], the Bose-Einstein condensation of charged bosons [30], as well as the mixed symmetry order parameters [31]. For example, the observed upward curvature in $Na_{0.35}CoO_2 \cdot yH_2O$ has been considered to result from the transition between two different pairing symmetries ($s + d$) that occur on different energy bands and gives the best fit to the experimental data [32].

On the other hand, we noted also that a similar behavior observed in some multiband systems such as $MgB_2$ [33] and iron-based superconductors [34], and can be explained well within a two-band scenario with taking into account the inter- and intra-band scattering [35]. In our case, the best fitted $H_{C2}(T)$ curves are shown in Fig. 2(d,e) and can be seen to agree very well with the experimental data [more details in Supporting Information], which seems to suggest the multi-band nature of the superconductivity. Within this theoretical framework and taking the 99% criterion, the $\mu_0H_{C2}(0)$ along the *c*-axis and the *ab*-plane were calculated to be ~23 T and 41 T, respectively. Both the estimated upper critical fields at zero temperature far exceeds the Pauli pair-breaking limit (given by $\mu_0H_P = 1.84\ T_C \approx 11.6$ T for $T_C = 6.3$ K) in a BCS weak-coupling case [36-37]. When employing the 90% criterion, the $\mu_0H_{C2}^{c}(0)$ and $\mu_0H_{C2}^{ab}(0)$ were estimated to be ~9 T and 31 T, respectively, and the corresponding anisotropic ratio $\gamma = \dfrac{H_{C2}^{c}(0)}{H_{C2}^{ab}(0)}$ in $Na_2CoSe_2O$ is ~ 3.3, which is a little higher than that of $Sr_{1-x}K_xFe_2As_2$ with $\gamma = 2$ [38], while it is much lower than high $T_C$ cuprates, for example $\gamma = 7 - 10$ for YBCO [39]. The high in-plane $\mu_0H_{C2}(0)$, exceeding the Pauli limiting field by a factor of 3 - 4, suggests a more exotic pairing mechanism in $Na_2CoSe_2O$. It is interesting that the upper critical field is far beyond that of $Na_{0.35}CoO_2 \cdot 1.3H_2O$ ($\mu_0H_{C2}(0)$ ~ 8 T for $\mu_0H\ //\ ab$) [40]. Nevertheless, a topologically non-trivial superconducting pairing via the nearest neighbor antiferromagnetic super-exchange in such triangular lattice has been proposed for $Na_xCoO_2 \cdot yH_2O$ [9], a similar mechanism may also be involved in $Na_2CoSe_2O$ due to the same geometric frustration, which warrant further investigation.

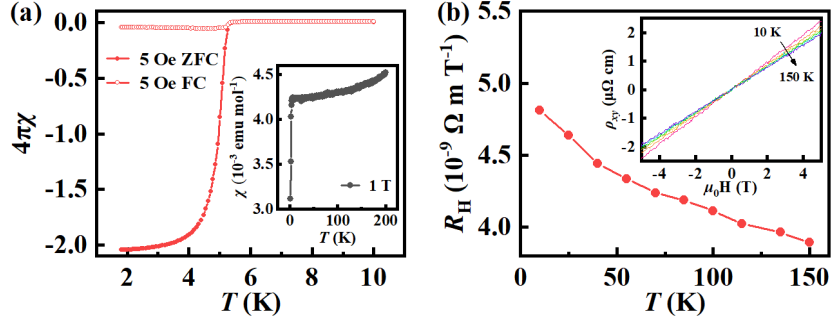

**Fig. 3. Magnetic susceptibility and Hall coefficient of $Na_2CoSe_2O$.** (a) Temperature dependence of the ZFC (filled circles) and FC (open circles) magnetic susceptibility for $Na_2CoSe_2O$ at a magnetic field of 5 Oe. The inset is the temperature dependence of the magnetic susceptibility $\chi(T)$ obtained in a magnetic field of 1 T for $\mu_0H \parallel c$. (b) Temperature dependence of Hall coefficient $R_H$ between 10 and 150 K. Inset: Magnetic field dependence of the Hall resistivity $\rho_{xy}$ at several temperatures for $Na_2CoSe_2O$.

In order to check whether the observed SC is of a bulk nature, we performed dc magnetic susceptibility $\chi(T)$ measurements in both zero-field cooling (ZFC) and field cooling (FC) modes under a magnetic field of 5 Oe, which reveals a superconducting transition close to 5.4 K, as shown in Fig. 3(a), roughly corresponding to the zero-resistivity temperature in Fig. 2(b). The difference between ZFC and FC strongly suggests the type II superconducting nature. The superconducting volume fraction estimated from the ZFC magnetization at 2 K was much larger than 100 %, without considering the demagnetization factor due to the irregular shape of the samples, which unambiguously proves the bulk superconductivity of the $Na_2CoSe_2O$ sample. The inset of Fig. 2(a) shows the normal-state magnetic susceptibility as a function of temperature, $\chi(T)$, for the $Na_2CoSe_2O$ crystals with a magnetic field of 1 T applied along the $c$-axis. Because the crystal is too small, several pieces are used, which might cause some uncertainty in the analysis of the magnetization. A drop in susceptibility at about 5 K corresponds to the superconducting transition. Noted that $\chi(T)$ increases monotonically with increasing temperature above $T_C$, seeming to indicate the existence of AF spin fluctuation in the normal state. Similar $\chi$-$T$ curve was also observed in Na$_{0.35}$CoO$_2$·1.3H$_2$O, in which a significant increase with decreasing $T$ at low temperatures was considered to be extrinsic [41]. At present stage, the nature of magnetic interactions in Na$_2$CoSe$_2$O is not clear yet and will be the subject of further experimental investigations, which may play a crucial role in the formation of superconducting states in such system, as discussed above.

To get more information about the conducting carriers, we measured the Hall coefficient in the normal state. From the $\rho_{xy}(H)$ data (inset of Fig. 3b), the Hall coefficient $R_H$ is determined through $R_H = \rho_{xy}/\mu_0 H$ and is plotted in Fig. 3(b), which shows a strong temperature dependent behavior. For a metallic material, the $T$-dependent $R_H$ is often explained by the multiband effect [42]. Indeed, according to the DFT calculations for Na$_2$CoSe$_2$O, five bands cross the Fermi level $E_F$ forming five hole pockets. Therefore, it is possible that the $T$-dependent $R_H$ comes from the multiband effect. Alternatively, the $T$-dependence of $R_H$ could also be caused by a magnetic skew scattering mechanism [43], which has been observed in various materials with the presence of magnetic moments. As the compound Na$_2$CoSe$_2$O contains Co element, and the Co valence state may be mainly Co IV, then the skew scattering mechanism might also work here. So here the strong $T$-dependent $R_H$ tells us that either the multiband effect or the skew scattering might be involved in the transport process in this material. Actually, large $T$-dependence of $R_H$ was also observed in the iron-based and copper-oxide superconductors, and was regarded as one of the exotic properties [44-45]. The positive $R_H$ over the whole temperature measured up to 150 K suggests that holes are the dominant carriers in transport, consistent with the results of our band calculations. A rough estimation based on the relation $R_H = 1/ne$ indicates that the carrier density is rather low, for example, at 10 K, the carrier density is about $1.3 \times 10^{21}$ cm$^{-3}$, which is similar to those of the iron-based superconductors and high-$T_C$ cuprates [44-45]. Thus Na$_2$CoSe$_2$O belongs to a class of poor conductors in the normal state.

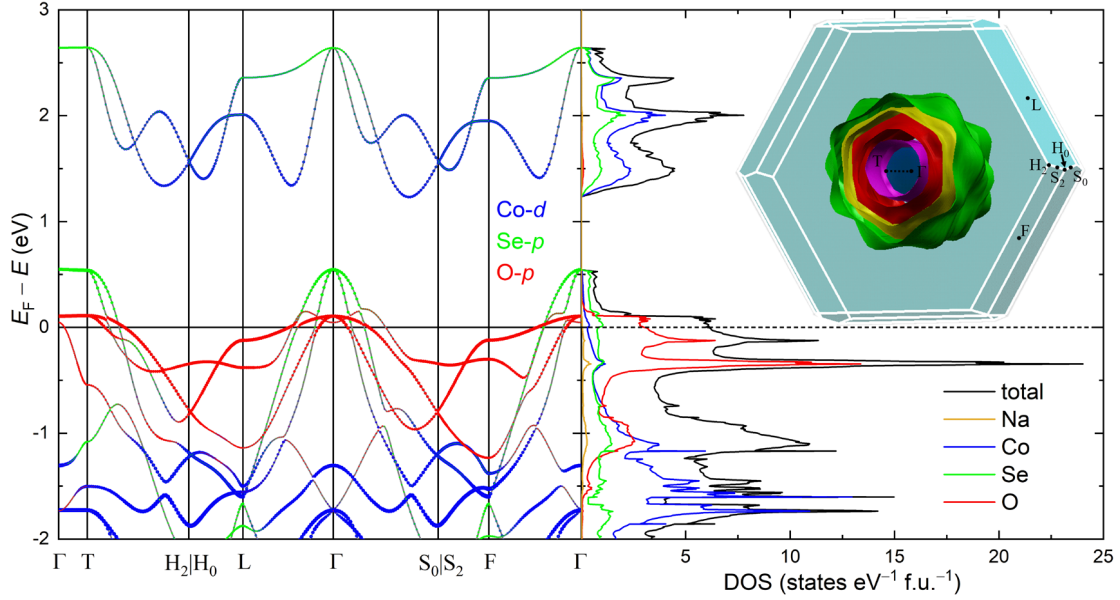

**Fig. 4. The band structure and Fermi surfaces of $Na_2CoSe_2O$ by DFT+ $U$ calculations.** Electronic band structure (left) and density of states (right) of $Na_2CoSe_2O$. Different colors represent contributions from different atoms/orbitals. Inset shows the Fermi surfaces of $Na_2CoSe_2O$ in the first Brillouin zone, in which the high symmetry points are labeled.

Fig. 4 shows the results of DFT + $U$ calculations. An indirect energy gap of about 0.8 eV can be identified. The $E_F$ locates near the top of the valence bands, implying $Na_2CoSe_2O$ to be a p-type semiconductor. There are multiple bands crossing $E_F$, consistent with the Hall/$H_{C2}$ results. Interestingly, states on $E_F$ are mainly contributed by the O-$2p$ and Se-$4p$ orbitals, while the Co-$3d$ states dominate below –1 eV. $Na_2CoSe_2O$ is thus a negative charge transfer material, which is as expected from the nominal oxidation state ($Co^{4+}$) of Co. With Co in such a high oxidation state, electrons tend to transfer from the ligand (Se) to Co, leaving additional holes in the ligand orbitals. Similar phenomena have been observed in materials with highly oxidized transition metals, such as $NaCuO_2$ [46], $NdNiO_3$ [47], and $CaFeO_3$ [48]. We also note that, while most of the negative charge transfer materials are insulators, $Na_2CoSe_2O$ serves as a rare example of negative charge transfer superconductor. The Bader analysis gives a valence configuration of $[Na_2O]^{0.43+}[CoSe_2]^{0.43-}$. In other words, electrons from the [$Na_2O$] layers are transferred into the [$CoSe_2$] layers, lowering the valence of

Co. The Co valence in $Na_2CoSe_2O$ is then estimated to be 3.57, which is comparable to the value of 3.46 for the superconducting $Na_{0.36}CoO_2 \cdot 1.3H_2O$ [49]. The DFT calculation suggests nearly-full Co $(t_{2g})^6$ orbitals, while the $e_g$ orbitals are empty. In other words, a low-spin $(t_{2g})^6(e_g)^0$ configuration is favored. We note that although the DFT + $U$ calculation suggests a low-spin $(t_{2g})^6$ configuration for Co, the actual spin configuration could still be $(t_{2g})^4(e_g)^2$, given that $Se^{2-}$ is a relatively weak ligand. Note that these calculations are nonmagnetic. A future neutron diffraction study will help to determine the actual spin configuration of Co. Moreover, when the Hubbard $U$ term is ignored, a much higher DOS for Co can be expected (see Figure S2). In both cases (with and without $U$), there are considerable amount of holes on the O and Se orbitals, resulting from the aforementioned charge transfer mechanism. This means that both the [$Na_2O$] and [$CoSe_2$] layers may be conducting. However, the hole mobility in these two layers can be very different. Note the O bands are nearly flat near the Fermi level (Figure 4), while the Se bands are steeper. This means that the holes in the [$Na_2O$] layers are heavier. The concentration of holes on ligands Se and O is as expected for negative charge-transfer (NCT) compounds, similar scenarios have been proposed in other NCT compounds such as $NdNiO_3$ [47] and pressurized cuprates [50]. The Fermi surfaces of $Na_2CoSe_2O$ consist of five sheets (see the inset of Fig. 4). All of them are hole-like surfaces surrounding the Γ point. Most of the Fermi surfaces, except for the innermost one, are quasi-2D, showing little energy dispersion along the Γ–T path. This feature is in line with the quasi-2D nature of the crystal structure.

Clearly, $Na_2CoSe_2O$ shares many aspects with the cuprates and Fe-based superconductors, such as two-dimensionality, low carrier concentration, multiple bands, very high upper critical fields, and also some hints of the strong electronic correlations and AF spin fluctuation, all of which point towards an exotic superconductivity in $Na_2CoSe_2O$. One difference is that, unlike the cuprates or iron-based superconductors, where the parent materials are AF Mott insulators or AF semimetals; $Na_2CoSe_2O$ shows superconductivity without explicit doping, where the charge transfer acts as a self-doping effect, which, coupled with the geometrical frustration, prevents the formation of long range magnetic ordering, and eventually leads to superconductivity. It has been well-established that the 2D [$CuO_2$] or [$Fe_2P_{C2}$] conducting layers modulated by the blocking layers (serve as charge reservoir) is critical for the high-$T_C$ superconductivity, in which various types of blocking layers have been realized, and several of them achieved high $T_C$. Similarly, since the [$CoSe_2$] layer is the key structural unit in such Co-based superconductor, analogous superconductors could be designed by modifying the blocking layers of [$Na_2O$]. Finding new components for the blocking layers is the key to discovering new superconductors with higher $T_C$. Additionally, building up multilayered cobalt chalcogenides may also lead to higher $T_C$ as the case of cuprates [51]. Multilayered iron-based superconductors with square lattices are still not discovered, but multilayered systems with triangular lattices have been achieved in iron sulfide (Smythite, $Fe_3S_4$) and chromium selenide $Na_{0.7}Cr_{2.3}Se_4$ [52-53]. Certainly, there are many other candidates with 3$d$ transition-metal-based 2D triangular lattices, which have similar chalcogenide chemistry to cobalt. This layered structure provides new hope and opportunity for exploring the exotic superconductivity in TMDs.

Finally, we would like to point out that much attentions have been focused in recent years on the 3$d$ transition-metal oxychalcogenides with square lattices, in which the possibility of high-$T_C$ superconductivity was theoretically predicted upon doping the Mott insulating state with electrons or holes [54-55]. Nevertheless, superconductivity has not yet been observed in such oxychalcogenides by now. Our result demonstrates for the first time, to our knowledge, that Co-based oxychalcogenide is indeed superconducting, but with a triangular lattice. Compared to the square lattice in iron-based superconductors, triangular lattice may lead to unconventional ground state and/or form geometric frustration in magnetic system, which has aroused great interest in the field of strongly correlated electron systems. For example, in addition to the superconductivity in $Na_xCoO_2 \cdot yH_2O$, the discovery of the heavy fermion behavior in the pyrochlore compound $LiV_2O_4$ [56-57], as well as the newly discovered Kagome lattice superconductors $AV_3Sb_5$ ($A$ = K, Rb, Cs) [58-59], have stimulated intensive studies of frustrated electron systems. There is no doubt that, the triangular lattice in Na$_2$CoSe$_2$O offers exciting playground and challenges for superconductivity, and there is still high potential to study and discover novel type of superconductor in such triangular system.

**CONCLUSIONS**

In conclusion, we have discovered a new layered oxychalcogenide superconductor. Its crystal structure, built by alternate stacking of [CoSe$_2$] conducting layers and [Na$_2$O] blocking layers, is similar to those of layered cuprates and Fe-based family. One difference is that, instead of the well-defined square lattices in Cu-O and Fe-As plane, Co atoms form a two-dimensional triangular lattice in Na$_2$CoSe$_2$O, in which superconducting state originates from the geometric frustrated [CoSe$_2$] layers. The preliminary physical property measurements indicated that this system also has many unique features, such as low carrier concentration, multi-band, and high upper critical fields, all of which point towards an exotic superconductivity in Na$_2$CoSe$_2$O. Based on the flexibility of the structure, new Co-based superconductors with various stacking structures can be designed by changing the blocking layers.

Our studies expand the categories of unconventional superconductors in 3$d$ transition-metal compounds, and open a door for finding new potential high-$T_C$ superconductors. It would be of interest in future studies to explore the similarities and differences between the layered cobalt oxychalcogenide and hydrated cobaltate from both theoretical and experimental perspectives, as well as infer something universal about unconventional superconductivity between these two-dimensional geometrically frustrated lattices and the well-defined square lattices of cuprates and iron-based family, which will help resolve their elusive superconducting mechanism.

## ASSOCIATED CONTENT

**Supporting Information**

Detailed crystal structure analysis of $Na_2CoSe_2O$, EDX measurements, Thermodynamic stability and HSAB rule analysis of $Na_2CoSe_2O$, Two-band model analysis of $H_{C2}(T)$ in two directions and DFT results without $U$.

## AUTHOR INFORMATION

**Corresponding Author**

*Corresponding author. E-mail: gfchen@iphy.ac.cn

**Author Contributions**

§J. Cheng and J. Bai contributed equally to this work.

**Notes**

The authors declare that they have no competing interests.

## ACKNOWLEDGMENTS

This work is supported by the Ministry of Science and Technology of China (Grant No. 2022YFA1403903), the National Natural Science Foundation of China (Grant No. 12274440), the Strategic Priority Research Program (B) of Chinese Academy of Sciences (Grant No. XDB33010100), and the Synergetic Extreme Condition User Facility (SECUF).